# An Adaptive Port Technique for Synthesising Rotational Components in Component Modal Synthesis Approaches

Xiang ZHAO[*], My Ha DAO

Institute of High Performance Computing (IHPC),
Agency for Science, Technology and Research (A*STAR), 1 Fusionopolis Way, #16-16 Connexis, Singapore 138632

## Abstract

Component Modal Synthesis (CMS) is a reduced order modelling method widely used for large-scale complex systems. It can effectively approximate system-level models through component synthesis, in which the repetitive geometrical components are modelled once and synthesised together. However, the conventional CMS only applies to systems with stationary components connected by strictly compatible ports, limiting it from modelling systems with moving components. This paper presents an adaptive port (AP) technique to extend CMS approaches for modelling parametric systems with rotational parts. To demonstrate the capability of the AP technique, we apply it to the Static Condensation Reduced Basis Element (SCRBE), one widely used variant of CMS approaches. The AP-based SCRBE (AP-SCRBE) can enforce the synthesis of rotational-stationary components over a shared adaptive port when the connecting surfaces of two components are discretisation-wise incompatible, which happens when one component moves relative to the others. Numerical experiments on the NREL 5MW wind turbine show that, in the context of rotational-stationary component synthesis, the AP-SCRBE can accurately and efficiently model the rotating rotor with pitch rotation of blades. It can produce almost identical results to a high-fidelity finite element model at two to three orders faster speeds.

---

[*] Corresponding author, E-mail: zhao_xiang@ihpc.a-star.edu.sg

# 1 Introduction

In modern engineering applications, high-fidelity numerical methods such as Finite Element (FE), Finite Difference (FD) and Finite Volume (FV) methods play an essential role in the numerical simulations of parametric systems governed by partial differential equations (PDEs). However, the computational overhead using the classical high-fidelity methods overgrows when dealing with large-scale complex systems, making the numerical simulations of those systems prolonged or practically prohibited even on high-performance computers. The unaffordable computational costs have limited the applications of high-fidelity methods in many time-critical and multi-query problems, often found in the context of design, optimisation and control.

Compared to the high-fidelity numerical methods, Reduced Order Modelling (ROM) or Model Order Reduction (MOR) methods have much lower computational costs by projecting the high-dimensional numerical discretisation of large-scale systems onto low-dimensional spaces. They are especially suitable for multi-query problems, where a family of similar problems in a particular parameter space is considered. Therefore, ROM/MOR methods are often employed as alternative approaches to speed up the multiple numerical simulations of parametric problems. The Proper Orthogonal Decomposition (POD) based methods [1-3] are the most well-known approaches for building ROMs. They rely on the singular value decomposition of a set of samples, which must be sufficiently large to represent the dominant low dimensional dynamics. Moreover, POD-based methods are usually constructed for a fixed domain or geometry and must be rebuilt once there are any geometrical variations. The Reduced Basis (RB) method [4-6] is another popular method which approximates high-fidelity solutions with fewer reduced basis/basis functions. The reduced basis can be pre-computed at an offline stage, while the approximation using the RB method happens at an online stage at a very low computational cost. The RB method allows variations in physical parameters, including geometrical parameters. There is much subsequent development and refinement to improve the accuracy of the RB method. By using the online enrichment [7-11], the quality of the reduced solution at the online stage is significantly enhanced, given the insufficient reduced basis generated at the offline stage. For example, Ohlberger and F. Schindler [9] use an online error estimator to adaptively enrich the solution space for parametric multi-scale problems. For nonlinear systems, Peherstorfer and Willcox [10] use a few selected new data from full order systems to build a reduced system that accurately approximates unseen scenarios.

For large parametric systems in engineering applications, the RB method is further extended under the framework of Component Modal Synthesis (CMS) [12-16]. The typical physical assets in engineering applications naturally have component synthesis features, where the geometries of the assets often comprise identical or similar components. For example, buildings, bridges, and gas/wind turbines are made of repeated sub-structures such as rooms, frames, and blades. Under the CMS framework, model reductions are performed for a few representative archetype components at an intradomain level. Then, the required components are instantiated and synthesised at an interdomain level to perform component-to-system level modelling for any parametric systems formed by the archetypes. The Multiscale RB method [17], the Multiscale FE method [18] and the Static Condensation Reduced Basis Element (SCRBE) [19-21] are examples of CMS approaches. However, the conventional CMS approaches only apply to systems with stationary components connected by strictly compatible ports, limiting them from modelling systems with moving components, such as an operational wind turbine. This work presents an adaptive port (AP) technique to relax this limitation in the CMS. Without loss of generality, we use the SCRBE method as an example of CMS approaches to demonstrate the capability of the AP technique, since the SCRBE method offers more flexibility for heterogenous components and has more rigorous system–level error bounds.

The SCRBE method originally developed by Huynh et al. [19-21] is a CMS version of the RB method [4] for modelling large-scale complex systems with many repetitive geometrical components. More specifically, the RB method is employed on individual components at an intradomain level, while the static condensation method [22] is applied at an interdomain level to synthesise the components into a system-level model. The components used in this computational architecture are interoperable, meaning that one archetype component can be connected to several other archetype components via ports or interfaces. For this reason, the SCRBE is a "bottom-up" computational strategy which can solve a family of problems related to possible combinations of the archetype components in the offline library [23]. In the last decade, the SCRBE method has gained growing popularity and has been successfully applied to linear elasticity [19-21, 24-26], heat transfer [27], acoustics [28] and lubrication [29]. The SCRBE method has also been combined with a Full Order Model (FOM) to address local nonlinearities. Ballani et al. [30] use a hybrid ROM-FOM approach, where the SCRBE method approximates the predominant part associated with linear elasticity, and a full FE model approximates the rest of the local nonlinearities. Alternatively, Zhang et al. [31] have developed a ROM approach, which employs the intrusive SCRBE to solve the linear subdomain and the non-intrusive Gaussian processes regression (GPR) to solve local nonlinearities.

Like other CMS approaches, the original SCRBE method [19-21] requires relatively strict conditions for component synthesis via pre-defined ports or interfaces. First, the shared port between two components must be geometrically compatible. Most importantly, the underlying computational grids of the two components must match exactly or be discretisationally compatible on the connecting port surface. Even a tiny difference in the grid topology or shape at the shared port will cause the component synthesis to fail. The latter condition prevents the SCRBE from being applied to modelling rotational components that move relative to each other. These problems are commonly seen in many engineering applications with moving parts.

There are a few methods that can address the incompatibility issue. The interface reduction method with constraint modes [32-34] is able to couple components with incompatible ports through approximation techniques. For example, the weak interface reduction method [32] enforces port compatibility by minimizing the compatibility error between connecting substructure interfaces. Alternatively, the mortar finite element method [35] allows the coupling of different discretisations across sub-regional boundaries, often used in full order FE models. However, there are very few papers in the literature on integrating the mortar method into the component-based ROM. This paper proposes an adaptive port (AP) technique to synthesise rotational and stationary components. It removes the port incompatibility during component synthesising by introducing an adaptive port with a dynamically generated buffer component.

The AP-enhanced Static Condensation Reduced Basis Element (AP-SCRBE) method allows the component synthesis over a shared adaptive port when the connecting surfaces of two components are discretisation-wise incompatible (e.g. having different port grids) but geometrically compatible (i.e. having the same port shape). It can even allow the component synthesis if the connecting surfaces have slightly different shapes. This ability enables the AP-SCRBE to model rotational components in a complex system, where the axial symmetric connecting surfaces of rotational components are consistently geometrically compatible but discretisation-wise incompatible. The technique is robust and computationally efficient, and it can be easily implemented without significantly affecting the efficiency of the original SCRBE method. For a demonstration, we use the National Renewable Energy Laboratory (NREL) 5MW wind turbine [36] with parameter variations to consider the rotation of turbine blades and rotor.

The rest of this paper is organised as follows: Section 2 briefly introduces the methodology of the CMS framework and the original SCRBE method. Section 3 proposes the AP-SCRBE method, which improves the SCRBE method by enabling the synthesis of both rotational and stationary components. Section 4 presents the results and analyses of the performance of the AP-SCRBE in a parametric study of a wind turbine example with both rotational and stationary components. Conclusions are drawn in Section 5.

## 2  A brief review of component modal synthesis framework

For completeness and conciseness, the CMS framework and the original SCRBE [19-21] method is first briefly reviewed to explain some basic ideas of the component-based approach.

### 2.1  Domain decomposition and component modal synthesis

The CMS is component-based framework, which relies on a library of pre-defined computational sub-domains or archetype components. In this library, the computational domain for the $m$-th archetype is defined as

$$\hat{\Omega}_m(\hat{\mu}_m), \qquad 1 < m < M \tag{1}$$

where $\hat{\mu}_m$ is the parameter vector associated with the $m$-th archetype, $M$ is the total number of archetypes. The $j$-th port on the $m$-th archetype is defined as

$$\hat{\gamma}_{m,j}(\hat{\mu}_m), \qquad 1 < j < n_m^\gamma \tag{2}$$

where $n_m^\gamma$ is the total number of available ports on the $m$-th archetype. The symbol ˆ in Eqs. (1) and (2) indicates that the definitions refer to an archetype rather than an instance of an archetype.

Before doing component synthesis, one should select the components needed for a particular problem and instantiate the components from the corresponding archetypes in the library. One can define

$$\Omega_i(\mu_i) = \hat{\Omega}_{\mathcal{M}(i)}(\hat{\mu}_i), \qquad 1 < i < I \tag{3}$$

where $I$ is the total number of instantiated components, and $\mathcal{M}$ maps the $i$-th instantiated component to the corresponding $\mathcal{M}(i)$-th archetype. Similarly, the instantiated ports for the $i$-th instantiated component are defined as

$$\gamma_{i,j}(\mu_i) = \hat{\gamma}_{\mathcal{M}(i),j}(\mu_i), \qquad 1 < j < n^{\gamma}_{\mathcal{M}(i)} \tag{4}$$

Once the selected components are synthesised together to form a system, the computational domain for the synthesised system is the union of the instantiated components

$$\overline{\Omega}(\mu) = \bigcup_{i=1}^{I} \overline{\Omega}_i(\mu_i), \tag{5}$$

where the bar sign $\overline{\Omega}$ indicates a close domain, including its boundary $\partial\Omega$. The instantiated components are connected by global ports $p$, $1 < p < n^{\Gamma}$, where $n^{\Gamma}$ is the total number of global ports. Each global port $p$ corresponds to an index set $\pi_p$, which is either the coincidence of two local ports of two instantiated components,

$$\pi_p = \{(i,j), (i',j')\} \tag{6}$$

or one local port of one instantiated component,

$$\pi_p = \{(i,j)\} \tag{7}$$

According to Ess.(6) and (7), one can also define a mapping from local ports to global ports,

$$p = \mathcal{G}_i(j) = \mathcal{G}_{i'}(j') \tag{8}$$

and an inverse mapping,

$$\begin{cases} j = \mathcal{G}_i^{-1}(p) \\ j' = \mathcal{G}_{i'}^{-1}(p) \end{cases} \tag{9}$$

Under geometric compatibility requirements, the instantiated components must not intersect in the component synthesis, and two instantiated components can only be connected by global port $\Gamma_p = \overline{\Omega}_i \cap \overline{\Omega}_j \neq \emptyset$. The global port connections should remain fixed for all parameter variations.

## 2.2 Governing equations

Without loss of generality, we consider a Partial Differential Equation (PDE) in a bilinear form defined in global function space $X(\Omega)$,

$$a(w, v; \mu) = f(v; \mu), \quad \forall v \in X(\Omega). \tag{10}$$

After the domain decomposition, the PDE in Eq.(10) can be re-written in a bilinear form in terms of the archetype components introduced in Section 2.1 , which is the summation of the PDEs for each component $i$

$$a(w, v; \mu) = \sum_{i=1}^{I} \hat{a}_{\mathcal{M}(i)}\big(w|_{\Omega_i(\mu_i)}, v|_{\Omega_i(\mu_i)}; \mu_i\big). \tag{11}$$

Similarly, the source term can also be decomposed and written in a bilinear form as

$$f(v;\mu) = \sum_{i=1}^{I} \hat{f}_{\mathcal{M}(i)}(v|_{\Omega_i(\mu_i)}; \mu_i), \tag{12}$$

To solve Eq.(10) numerically, the Finite Element (FE) discretisation is applied over the archetype component reference domains $\hat{\Omega}_m$. Then, the FE space for the entire discrete system $X^h(\Omega)$ becomes

$$X^h(\Omega) = \oplus_{i=1}^{I} X^h_{\mathcal{M}(i)}(\hat{\Omega}_m) \tag{13}$$

where $X^h(\Omega) \subset X(\Omega)$. The FE approximation to $u^h(\mu)$ should satisfy

$$a(u^h(\mu), v; \mu) = f(v; \mu), \quad \forall v \in X^h(\Omega). \tag{14}$$

On the ports of components, $P^h_{m,j}$ denotes the corresponding FE discretisation port spaces. Under the discretisation requirement, the global port made of two local ports should satisfy a conforming condition. This system-specific constraint for archetype components requires the meshes of two local ports in Eq.(6) to be exactly the same. Here, $X^h(\Omega)$ and $P^h_{m,j}$ are the discretisation spaces on which the CMS is applied. After being rotated, the meshes of two local ports become misaligned, making the component synthesising inapplicable. Therefore, the discretisation requirement limits the conventional CMS approaches from being applied to rotational components that can be geometrically compatible but discretisationally incompatible. We will discuss discretisation compatibility more later in Section 3.

## 2.3 Static Condensation Reduced Basis Element (SCRBE) method

The SCRBE method is a variant of CMS approaches, in which the RB method is applied at the intradomain model reduction. Before introducing the SCRBE method, some spaces and basis functions are defined. First, the bubble space for each archetype component is defined as

$$B^h_{m;0} = \{v \in X^h_m : v|_{\hat{\gamma}_{m,j}} = 0, \quad 1 \leq j \leq n^\gamma_m\}, \quad 1 \leq m \leq M \tag{15}$$

and the interface functions for port space $P^h_{m,j}$ is

$$\chi_{m,j,k} \in P^h_{m,j}, \quad 1 \leq k \leq \mathcal{N}^\gamma_{m,j} \tag{16}$$

The discretisation compatibility requires

$$\chi_{\mathcal{M}(i),j,k} = \chi_{\mathcal{M}(i'),j',k}, \quad 1 \leq k \leq \mathcal{N}^\Gamma_p = \mathcal{N}^\Gamma_{\mathcal{M}(i),j} = \mathcal{N}^\Gamma_{\mathcal{M}(i'),j'} \tag{17}$$

for each global port made of two local ports as defined in Eq.(6). The construction of the interface functions $\chi_{m,j,k}$ is essential for the final performance of the synthesised system. The interface functions $\chi_{m,j,k}$ can be computed by the eigen expansion [19], the port reduction technique [37, 38], the local approximation spaces [39] and the oversampling method [40]. The port reduction technique is used in this work since it has a relatively faster convergence rate and is easy to implement. The interface functions have a lifting effect on the components connected to a global port. One then uses $\psi_{m,j,k}$ to denote the extension functions of the interface functions $\chi_{m,j,k}$ into archetype component $\hat{\Omega}_m$ due to the lifting effect. The extension functions $\psi_{m,j,k}$ could be economically computed using the discrete harmonic extensions [41].

The bubble functions, $b^h_{i,j,k} \in B^h_{\mathcal{M}(i);0}$, should satisfy

$$a(b^h_{i,j,k}, v; \mu_i) = -a(\psi_{\mathcal{M}(i),j,k}, v; \mu_i), \quad \forall v \in B^h_{\mathcal{M}(i);0}. \tag{18}$$

Besides that, one should also consider the bubble functions due to external force, called force-induced bubble functions $b^{f;h}_{i,j,k}$. They should satisfy

$$a(b^{f;h}_i, v; \mu_i) = f_{\mathcal{M}(i)}(v; \mu_i), \quad \forall v \in B^h_{\mathcal{M}(i);0}. \tag{19}$$

On each component $\Omega_i$, the SCRBE method approximate the solution of Eq. (14) in terms of

$$u^h(\mu)|_{\Omega_i} = b^{f;h}_i(\mu_i) + \sum_{j=1}^{n^\gamma_{\mathcal{M}(i)}} \sum_{k=1}^{\mathcal{N}^\gamma_{\mathcal{M}(i),j}} \mathbb{U}_{\mathcal{G}_i(j),k} \phi^h_{i,j,k}(\mu_i) \tag{20}$$

where

$$\phi_{i,j,k} \equiv b^h_{i,j,k} + \psi_{\mathcal{M}(i),j,k} \tag{21}$$

and $\mathbb{U}_{\mathcal{G}_i(j),k}$ are the coefficients to be determined. One can sum Eq. (20) over all components to obtain the approximated global solution of Eq. (14), which has a form of

$$u^h(\mu) = \sum_{i=1}^{I} b^{f;h}_i(\mu_i) + \sum_{p=1}^{n^\Gamma} \sum_{k=1}^{\mathcal{N}^\Gamma_p} \mathbb{U}_{p,k} \Phi_{p,k} \tag{22}$$

where $\Phi_{p,k}$ are interface functions,

$$\Phi_{p,k} \equiv \sum_{(i,j) \in \pi_p} \phi_{i,j,k} \tag{23}$$

and $\mathbb{U}_{p,k}, 1 \leq k \leq \mathcal{N}^\Gamma_p, 1 \leq p \leq n^\Gamma$ are the interface function coefficients to be determined. Inserting Eq. (22) into Eq. (14) gives

$$\sum_{p=1}^{n^\Gamma} \sum_{k=1}^{\mathcal{N}^\Gamma_p} \mathbb{U}_{p,k} a(\Phi_{p,k}, v; \mu) = f(v; \mu) - \sum_{i=1}^{I} a(b^{f;h}_i(\mu_i), v; \mu), \quad \forall v \in X^h(\Omega). \tag{24}$$

Eqs. (18)-(24) conclude the static condensation (SC). After the SC, the unknowns in the system are reduced to the coefficients $\mathbb{U}_{p,k}$. Thus, the total number of unknowns is reduced to $n_{\mathrm{SC}} \equiv \sum_{p=1}^{n^\Gamma} \mathcal{N}^\Gamma_p$ from $N_{\mathrm{total}} = \dim(X^h(\Omega))$, with $n_{\mathrm{SC}} \ll N_{\mathrm{total}}$.

The computation of the basis functions in the SC is very costly, especially when solving the bubble functions on components. In the SCRBE method, the heavy computation is significantly relaxed by introducing a Reduced Basis (RB) to approximate those bubble functions [5]. More specifically, instead of directly solving Eqs. (18) and (19), the bubble functions $b^h_{i,j,k}$ and $b^{f;h}_i$ are approximated by $\tilde{b}_{i,j,k}$ and $\tilde{b}^f_i \approx b^{f;h}_i$, which satisfy

$$a(\tilde{b}_{i,j,k}, v; \mu_i) = -a(\psi_{\mathcal{M}(i),j,k}, v; \mu_i), \quad \forall v \in \tilde{B}_{\mathcal{M}(i),j,k;0} \tag{25}$$

and

$$a(\tilde{b}_i^f, v; \mu_i) = f_{\mathcal{M}(i)}(v; \mu_i), \quad \forall v \in \tilde{B}_{\mathcal{M}(i);0}^f. \tag{26}$$

The RB solution spaces $\tilde{B}_{\mathcal{M}(i),j,k;0}$ and $\tilde{B}_{\mathcal{M}(i);0}^f$ of Eqs. (25) and (26) are constructed by the Greedy algorithm [42]. Then, $\phi_{i,j,k}$ and $\Phi_{p,k}$ in Eqs. (21) and (23) are replaced by the RB approximations $\tilde{\phi}_{i,j,k}$ and $\widetilde{\Phi}_{p,k}$

$$\tilde{\phi}_{i,j,k} \equiv \tilde{b}_{i,j,k} + \psi_{\mathcal{M}(i),j,k} \tag{27}$$

and

$$\widetilde{\Phi}_{p,k} \equiv \sum_{(i,j) \in \pi_p} \tilde{\phi}_{i,j,k} \tag{28}$$

With the RB approximations, the corresponding SCRBE system of dimension $n_{SC}$ can be constructed in the form of

$$\widetilde{\mathbb{A}}(\mu)\widetilde{\mathbb{U}}(\mu) = \widetilde{\mathbb{F}}(\mu) \tag{29}$$

where

$$\widetilde{\mathbb{A}}_{(p,k),(p',k')}(\mu) \equiv a(\widetilde{\Phi}_{p,k}(\mu), \widetilde{\Phi}_{p',k'}(\mu); \mu) \tag{30}$$

$$\widetilde{\mathbb{F}}_{(p',k')}(\mu) \equiv f(\widetilde{\Phi}_{p',k'}(\mu); \mu) - \sum_{i=1}^{I} a(\tilde{b}_i^f(\mu_i), \widetilde{\Phi}_{p',k'}(\mu); \mu) \tag{31}$$

and $\widetilde{\mathbb{U}}(\mu)$ is the vector of unknowns. The SCRBE method applies an offline-online computational paradigm. At the offline stage, the basis functions and RB approximation spaces for all archetype components are constructed. At the online stage, the archetype components are selected for a parametric problem and instantiated for assembling a system model very rapidly via the component synthesis, which has high flexibility in topological parameterisation and does not computationally scale with the dimension of FE spaces. Further details of the SCRBE method can be found in [19-21].

## 3 Adaptive Port Static Condensation Reduced Basis Element (AP-SCRBE) method

Having reviewed the original SCRBE method and its limitation in the component synthesis, we propose an enhanced version, namely the Adaptive Port Static Condensation Reduced Basis Element (AP-SCRBE) method. As discussed in Section 2.3, the AP-SCRBE method aims to lift the port discretisation compatibility constraints and efficiently compute the basis functions for rotational components. It allows rotational components to be included in the component synthesis and potentially other types of moving components.

Regarding the scope of applications, this method can be used to solve parametric problems where the rotational components in a complex system are at various rotation angles. This work applies the AP-SCRBE method to the elastic modelling of a wind turbine with parameter variations in its blades and rotor hub. In Section 4, the turbine blades are at different pitch angles from -90° to 90°; the hub is also rotated from 0° to 90°. Beyond this, this technique can also be applied to other systems with rotational components, such as propellers at different rotation angles and aircraft with wings at different pitch angles. In these applications, rotational components are synthesised with stationary components via connecting ports at different angles, which can be effectively modelled using the AP-SCRBE method.

### 3.1 Adaptive port technique

The core of the proposed AP-SCRBE method is the adaptive port technique to overcome the strict geometric and port discretisation compatibility requirements. As illustrated using a 2D example in Figure 1(a), two components can only be connected to form a global port 1 if component 1 and component 2 are geometrically and discretisationally compatible on the connecting interface. In other words, the connecting ports of the two components should have the same shape, and the underlying interface FE grids on the connecting ports must be identical. The original SCRBE method tries to satisfy the port discretisation compatibility by fixing the same interface grids for all possible connections of various archetype components via this global port [19]. This approach can guarantee the port discretisation compatibility for a fixed number of stationary components. However, it cannot handle parametric problems where some components consistently move due to the changes in orientation. For example, in Figure 1(b), if component 2 is flipped over, the grids on the connecting ports become misaligned. In more general 3D scenarios, the same problem can happen if one component is rotated by certain degrees about the horizontal axis, leading to incompatible port grids and, therefore, unconnectable components.

The adaptive port technique is proposed to overcome the discretisational incompatibility in port 1 (Figure 1(b)) for a parametric problem, where component 2 is rotated to an arbitrary angle. The technique is illustrated in Figure 1(c). Without loss of generality, we assume component 1 to be a fixed component and component 2 to be a rotational component about a horizontal axis. Component 2 consists of an adaptive buffer component $2^b$ and a main body component $2^*$. The buffer component $2^b$ is a relatively thin layer, while the main body component $2^*$ makes up the majority of component 2. When component 2 is rotated, the underlying FE grids (internal elements and boundary elements) of components 1 and $2^*$ remain unchanged as the pre-generated grids at the offline stage, while the internal FE grid of component $2^b$ is dynamically generated online and serves as the adaptor between component 1 and component $2^*$.

Two conditions should be satisfied for the adaptive port technique to work correctly and efficiently. First, the left and right port grids of component $2^b$ must be compatible with the right port grid of component 1 via port 1 and the left port grid of component $2^*$ via port 2, to guarantee discretisational compatibility. Second, since the basis functions $\tilde{b}_{\text{component2,port1}}$ and $\tilde{b}^f_{\text{component2,port1}}$ for component 2 are local to the neighbouring elements of port 1, they must be re-computed online at a low computational cost after the rotation of component $2^*$.

It is relatively easy to meet the first condition. The internal elements in buffer component $2^b$ are re-generated with boundary constraints to preserve the pre-defined port grid compatibility on ports 1 and 2. More specifically, the boundary constraints keep the left port grid of component $2^b$ identical to the right port grid of component 1, and the right port grid of component $2^b$ identical to the left port grid of component $2^*$, after component $2^*$ is rotated. Then, only the internal elements in component $2^b$ are re-generated, and the interface grids of the two ports are preserved. The element quality in buffer component $2^b$ can be controlled during the grid re-generation to ensure the stiffness matrix is well-conditioned. In practice, a method such as the constrained Delaunay tetrahedralization [43, 44] could be used for this purpose. Since component $2^b$ is a relatively thin layer of grids, the constrained Delaunay tetrahedralisation is computationally inexpensive at the online stage.

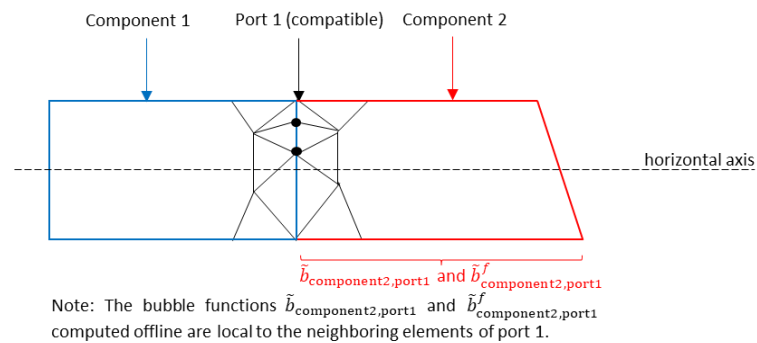

Note: The bubble functions $\tilde{b}_{component2,port1}$ and $\tilde{b}^f_{component2,port1}$ computed offline are local to the neighboring elements of port 1.

**(a) Compatible local ports**

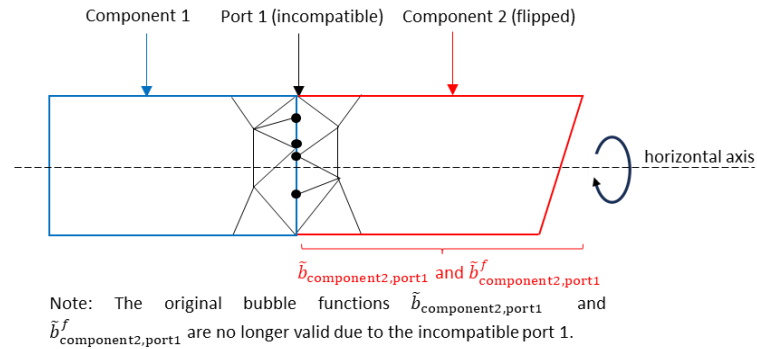

Note: The original bubble functions $\tilde{b}_{component2,port1}$ and $\tilde{b}^f_{component2,port1}$ are no longer valid due to the incompatible port 1.

**(b) Incompatible local ports after rotation**

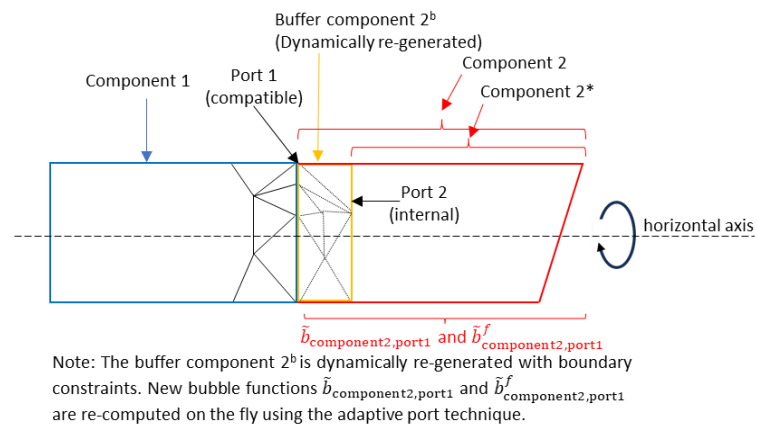

Note: The buffer component $2^b$ is dynamically re-generated with boundary constraints. New bubble functions $\tilde{b}_{component2,port1}$ and $\tilde{b}^f_{component2,port1}$ are re-computed on the fly using the adaptive port technique.

**(c) With Adaptive port**

Figure 1: Illustration of the adaptive port technology.

On the other hand, it takes relatively more effort to meet the second condition. At the online computation stage, component 2 is made of a dynamically generated component $2^b$ and a rotated component $2^*$. The computational efficiency for calculating the new basis functions for component 2 is especially critical, which will be discussed in Section 3.2 .

## 3.2 Basis function for rotational components

For an arbitrary rotational component $i$, one can now express $\tilde{b}_i$ in Eq. (25) (subscripts $j, k$ are omitted for conciseness) and $\tilde{b}_i^f$ in Eq. (26) in terms of

$$\tilde{b}_i = \sum_{l=1}^{N_i} u_{i,l}\zeta_{i,l} = \boldsymbol{u}_i^T \boldsymbol{\zeta}_i \qquad (32)$$

and

$$\tilde{b}_i^f = \sum_{l=1}^{N_i^f} u_{i,l}^f \zeta_l^f = \left(\boldsymbol{u}_i^f\right)^T \boldsymbol{\zeta}_i^f \qquad (33)$$

where $\boldsymbol{\zeta}_i$ and $\boldsymbol{\zeta}_i^f$ are the reduced basis obtained using the standard Greedy algorithm [42], $\boldsymbol{u}_i$ and $\boldsymbol{u}_i^f$ are the undetermined coefficients for $\boldsymbol{\zeta}_i$ and $\boldsymbol{\zeta}_i^f$, $N_i$ and $N_i^f$ are the total numbers of reduced basis used for approximating $\tilde{b}_i$ and $\tilde{b}_i^f$. Then, for component $i$, by substituting Eq. (32) into Eq. (25), and Eq. (33) into Eq. (26), one has

$$\sum_{l=1}^{N_i} u_{i,l} a(\zeta_{i,l}, v; \mu_i) = -a(\psi_{\mathcal{M}(i)}, v; \mu_i), \quad \forall v \in \tilde{B}_{\mathcal{M}(i);0} \qquad (34)$$

and

$$\sum_{l=1}^{N_i^f} u_{i,l}^f a(\zeta_{i,l}^f, v; \mu_i) = f_{\mathcal{M}(i)}(v; \mu_i), \quad \forall v \in \tilde{B}_{\mathcal{M}(i);0}^f. \qquad (35)$$

which can be expressed in matrix form as

$$\mathcal{A}\mathcal{U} = \mathcal{F} \qquad (36)$$

and

$$\mathcal{A}^f \mathcal{U}^f = \mathcal{F}^f \qquad (37)$$

where

$$\mathcal{A}_{l,l'} \equiv a(\zeta_{i,l}, \zeta_{i,l'}; \mu) \qquad (38)$$

$$\mathcal{A}_{l,l'}^f \equiv a\left(\zeta_{i,l}^f, \zeta_{i,l'}^f; \mu\right) \qquad (39)$$

$$\mathcal{F}_{l'} \equiv -a(\psi_{\mathcal{M}(i)}, \zeta_{i,l'}; \mu) \qquad (40)$$

$$\mathcal{F}_{l'}^{f} \equiv f_{\mathcal{M}(i)}\left(\zeta_{i,l'}^{f}; \mu\right) \tag{41}$$

$$\mathcal{U}_{l'} = \boldsymbol{u}_{i,l'} \tag{42}$$

$$\mathcal{U}_{l'}^{f} = \boldsymbol{u}_{i,l'}^{f} \tag{43}$$

Once $\mathcal{U}$ and $\mathcal{U}^f$ are computed by solving Eqs. (36) and (37), $\tilde{b}_i$ and $\tilde{b}_i^f$ can then be evaluated using Eqs. (32) and (33). The process can be expressed as

$$\tilde{b}_i = (\mathcal{A}^{-1}\mathcal{F})^T \boldsymbol{\zeta}_i \tag{44}$$

$$\tilde{b}_i^f = [(\mathcal{A}^f)^{-1}\mathcal{F}^f]^T \boldsymbol{\zeta}_i^f. \tag{45}$$

The evaluations of the basis functions $\tilde{b}_i$ and $\tilde{b}_i^f$ require inversions of $\mathcal{A}$ and $\mathcal{A}^f$ for each archetype component $i$, which is computationally relatively expensive. In the proposed adaptive port technique, if component $i$ is a stationary archetype component, the evaluation is performed offline like the original SCRBE method. On the other hand, if component $i$ is a rotational archetype component, $\tilde{b}_i$ and $\tilde{b}_i^f$ could be evaluated efficiently on the fly. Towards this goal, we apply the decomposition strategy for rotational components discussed earlier. A rotational component $i$ is decomposed into an adaptive buffer layer $i^b$ and its main body $i^m$. The buffer layer $i^b$ is generated on the fly, while the main body $i^m$ is pre-generated at the offline stage and is subject to rotational operations at the online stage. Therefore, one can decompose matrices $\mathcal{A}$ and $\mathcal{A}^f$ into

$$\mathcal{A} = \begin{bmatrix} \mathcal{R}(\theta)\mathcal{A}^{m,m}\mathcal{R}^T(\theta) & \mathcal{A}^{m,b} \\ \mathcal{A}^{b,m} & \mathcal{A}^{b,b} \end{bmatrix}, \tag{46}$$

$$\mathcal{A}^f = \begin{bmatrix} \mathcal{R}(\theta)\mathcal{A}^{f;m,m}\mathcal{R}^T(\theta) & \mathcal{A}^{f;m,b} \\ \mathcal{A}^{f;b,m} & \mathcal{A}^{f;b,b} \end{bmatrix}. \tag{47}$$

where $\mathcal{R}(\theta)$ is a rotation matrix acting on the main body $i^m$, $\theta$ is the rotation angle. $\mathcal{A}^{m,m}$ is pre-generated, corresponding to the underlying FE discretisation of the main body $i^m$ in the full FE space on $i^m$, while $\mathcal{A}^{b,b}$, $\mathcal{A}^{m,b}$ and $\mathcal{A}^{b,m}$ are generated on the fly in much smaller spaces as follows:

$$\begin{cases} \mathcal{A}, \mathcal{A}^f \in \mathbb{R}^{N_i \times N_i} \\ \mathcal{A}^{m,m}, \mathcal{A}^{f;m,m} \in \mathbb{R}^{N_{i^m} \times N_{i^m}} \\ \mathcal{A}^{m,b}, \mathcal{A}^{f;m,b} \in \mathbb{R}^{N_{i^m} \times N_{i^b}} \\ \mathcal{A}^{b,m}, \mathcal{A}^{f;b,m} \in \mathbb{R}^{N_{i^b} \times N_{i^m}} \\ \mathcal{A}^{b,b}, \mathcal{A}^{f;b,b} \in \mathbb{R}^{N_{i^b} \times N_{i^b}} \end{cases} \tag{48}$$

where

$$\begin{cases} N_i = \dim\left(X^h(\Omega_i)\right) \\ N_{i^m} = \dim\left(X^h(\Omega_{i^m})\right) \\ N_{i^b} = \dim\left(X^h(\Omega_{i^b})\right) \end{cases} \tag{49}$$

Since the buffer layer $i^b$ is relatively much thinner than the main body $i^m$, one has

$$N_{i^b} \ll N_{i^m} < N_i \tag{50}$$

In the original SCRBE, $[\mathcal{A}^{m,m}]^{-1}$ and $[\mathcal{A}^{f;m,m}]^{-1}$ are pre-computed for all archetype components in the library. However, for the rotational archetype component, the pre-computation is no longer applicable, because the rotation operator $\mathcal{R}(\theta)$ in Eqs. (46) and (47) is determined on the fly. To resolve this issue, the following identity for 2x2 block matrix inversion [45, 46] is applied,

$$\begin{bmatrix} A & B \\ C & D \end{bmatrix}^{-1} = \begin{bmatrix} A^{-1} + A^{-1}B(D - CA^{-1}B)^{-1}CA^{-1} & -A^{-1}B(D - CA^{-1}B)^{-1} \\ -(D - CA^{-1}B)^{-1}CA^{-1} & (D - CA^{-1}B)^{-1} \end{bmatrix} \tag{51}$$

which can be further written as

$$\begin{bmatrix} A & B \\ C & D \end{bmatrix}^{-1} = \begin{bmatrix} A^{-1} + H & -G \\ -G^T & F \end{bmatrix} \tag{52}$$

where

$$E = A^{-1}B \tag{53}$$
$$F = (D - CE)^{-1} \tag{54}$$
$$G = EF \tag{55}$$
$$H = GE^T \tag{56}$$

If one set

$$\mathcal{A} = \begin{bmatrix} A & B \\ C & D \end{bmatrix} \tag{57}$$

and

$$\begin{cases} A = \mathcal{R}(\theta)\mathcal{A}^{m,m}\mathcal{R}^T(\theta) \\ B = \mathcal{A}^{m,b} \\ C = \mathcal{A}^{b,m} \\ D = \mathcal{A}^{b,b} \end{cases} \tag{58}$$

Then, the inverse matrix $\mathcal{A}^{-1}$ for the rotational component can be computed by

$$\mathcal{A}^{-1} = \begin{bmatrix} \mathcal{R}(\theta)\mathcal{A}^{m,m}\mathcal{R}^T(\theta) & \mathcal{A}^{m,b} \\ \mathcal{A}^{b,m} & \mathcal{A}^{b,b} \end{bmatrix}^{-1}$$
$$= \begin{bmatrix} \mathcal{R}(\theta)[\mathcal{A}^{m,m}]^{-1}\mathcal{R}^T(\theta) + H & -G \\ -G^T & F \end{bmatrix} \tag{59}$$

where

$$E = \mathcal{R}(\theta)[\mathcal{A}^{m,m}]^{-1}\mathcal{R}^T(\theta)\mathcal{A}^{m,b} \in \mathbb{R}^{N_{i^m} \times N_{i^m}} \tag{60}$$

$$F = (\mathcal{A}^{b,b} - \mathcal{A}^{b,m}E)^{-1} \in \mathbb{R}^{N_{i^b} \times N_{i^b}} \tag{61}$$

$$G = EF \in \mathbb{R}^{N_{i^m} \times N_{i^b}} \tag{62}$$

$$H = GE^T \in \mathbb{R}^{N_{i^b} \times N_{i^b}} \tag{63}$$

Similarly, the inverse matrix $[\mathcal{A}^f]^{-1}$ for the rotational component can be computed by

$$
\begin{aligned}
[\mathcal{A}^f]^{-1} &= \begin{bmatrix} \mathcal{R}(\theta)\mathcal{A}^{f;m,m}\mathcal{R}^T(\theta) & \mathcal{A}^{f;m,b} \\ \mathcal{A}^{f;b,m} & \mathcal{A}^{f;b,b} \end{bmatrix}^{-1} \\
&= \begin{bmatrix} \mathcal{R}(\theta)[\mathcal{A}^{f;m,m}]^{-1}\mathcal{R}^T(\theta) + H^f & -G^f \\ -[G^f]^T & F^f \end{bmatrix}
\end{aligned}
\tag{64}
$$

where

$$E^f = \mathcal{R}(\theta)[\mathcal{A}^{f;m,m}]^{-1}\mathcal{R}^T(\theta)\mathcal{A}^{f;m,b} \in \mathbb{R}^{N_{i^m} \times N_{i^m}} \tag{65}$$

$$F^f = (\mathcal{A}^{f;b,b} - \mathcal{A}^{f;b,m}E^f)^{-1} \in \mathbb{R}^{N_{i^b} \times N_{i^b}} \tag{66}$$

$$G^f = E^f F^f \in \mathbb{R}^{N_{i^m} \times N_{i^b}} \tag{67}$$

$$H^f = G^f [E^f]^T \in \mathbb{R}^{N_{i^b} \times N_{i^b}} \tag{68}$$

One should note that $[\mathcal{A}^{m,m}]^{-1}$ and $[\mathcal{A}^{f;m,m}]^{-1}$ in Eqs. (59) and (64) are pre-computed at the offline stage. The online evaluation of $\mathcal{A}^{-1}$ and $[\mathcal{A}^f]^{-1}$ requires the additional evaluations of Eqs. (60) to (63), and Eqs. (65) to (68). In this evaluation procedure, only Eqs. (61) and (66) require the computation of matrix inversion in a very small space since $F, F^f \in \mathbb{R}^{N_{i^b} \times N_{i^b}}$ are small square matrices corresponding to the thin buffer layer $i^b$. The rest of the computations require multiplications of skinny sparse matrices and square sparse matrices, and therefore can also be done with trivial efforts. Basically, by applying the adaptive port technique, the expensive direct evaluation of $\mathcal{A}^{-1} \in \mathbb{R}^{N_i \times N_i}$ and $[\mathcal{A}^f]^{-1} \in \mathbb{R}^{N_i \times N_i}$ in the full discretisation space on $\Omega_i$ is transferred into a much smaller discretisation space of $\Omega_{i^b}$ with the help of the dynamically generated adaptive layer. One now can plug Eq.(59) into Eq.(44) and Eq.(64) into Eq.(45) to obtain basis functions $\tilde{b}_i$ and $\tilde{b}_i^f$ as follows,

$$
\begin{aligned}
\tilde{b}_i &= \left( \begin{bmatrix} \mathcal{R}(\theta)[\mathcal{A}^{m,m}]^{-1}\mathcal{R}^T(\theta) + H & -G \\ -G^T & F \end{bmatrix} \mathcal{F} \right)^T \zeta_i \\
&= \mathcal{F}^T \begin{bmatrix} \mathcal{R}(\theta)[\mathcal{A}^{m,m}]^{-1}\mathcal{R}^T(\theta) + H & -G \\ -G^T & F \end{bmatrix}^T \zeta_i
\end{aligned}
\tag{69}
$$

$$
\begin{aligned}
\tilde{b}_i^f &= \left[ \begin{bmatrix} \mathcal{R}(\theta)[\mathcal{A}^{f;m,m}]^{-1}\mathcal{R}^T(\theta) + H^f & -G^f \\ -[G^f]^T & F^f \end{bmatrix} \mathcal{F}^f \right]^T \zeta_i^f \\
&= (\mathcal{F}^f)^T \begin{bmatrix} \mathcal{R}(\theta)[\mathcal{A}^{f;m,m}]^{-1}\mathcal{R}^T(\theta) + H^f & -G^f \\ -[G^f]^T & F^f \end{bmatrix}^T \zeta_i
\end{aligned}
\tag{70}
$$

The proposed AP-SCRBE method extends the original SCRBE method to systems with rotational components in parametric studies. Although the traditional FE approach can also be used for the parametric problem by generating new meshes and computing solutions for possible configurations, the AP-SCRBE method can explore the solutions in the given parametric space much more rapidly. The breakdowns of computational time for the proposed AP-SCRBE method are listed in Table 1 in comparison with the high-fidelity FE approach. The online evaluation of the AP-SCRBE requires the online evaluation time to generate new grids for buffer layers dynamically. The FE approach can also use the same strategy to generate regional grids near rotational ports while keeping the grids of the other parts unchanged. Additionally, the AP-SCRBE needs to compute the new basis function for rotational components on the fly, which is essentially the procedure described by Eqs. (44), (45), (59) to (68). After that, a system is synthesised using the AP-SCRBE, which is computationally much faster than the FE approach. We will analyse the time consumption in more detail in Section 4.

Table 1: Computational time breakdowns of the high-fidelity FEM and the AP-SCRBE.

|   | AP-SCRBE | High-fidelity FEM |
|---|---|---|
| 1. | Time to generate new grid for buffer layer | Time to generate a new buffer grid |
| 2. | Time to solve ROM | Time to solve FEM |
| 3. | Time to compute new bubble functions | N.A. |

### 3.3 Algorithm

The algorithm of the AP-SCRBE method is summarized as follows,

**Offline:**

(1) Compute the reduced basis $\boldsymbol{\zeta}_i$ and $\boldsymbol{\zeta}_i^f$ in Eqs. (32) and (33) using the standard Greedy algorithm [42].

(2) Compute bubble functions $\tilde{b}_i$ and $\tilde{b}_i^f$ for all archetype components using Eqs. (44) and (45).

**Online:**

(1) If the instantiated component is rotational or has a non-conforming port, then re-compute bubble functions $\tilde{b}_i$ and $\tilde{b}_i^f$ on the fly using Eqs. (69) and (70), where $E$ to $H$ are computed using Eqs. (60) to (63), and $E^f$ to $H^f$ using Eqs. (65) to (68). Otherwise, skip this step.

(2) Solve the reduced system Eq. (29), which is formed using Eqs. (27), (28), (30) and (31) with the updated bubble functions $\tilde{b}_i$ and $\tilde{b}_i^f$.

## 4 Results and discussions

This section presents numerical results to demonstrate the capability of the AP-SCRBE method in modelling elastic systems with rotational components. In particular, a wind turbine system, with many repetitive sub-structures and rotational parts, is considered. It will be demonstrated that, by applying the AP-SCRBE method, one can quickly solve a family of related problems derived from topology and parameter variations. The numerical results are all based on the same offline archetype component library, which demonstrates the ability of the AP-SCRBE method to rapidly explore the solutions to parametric problems.

## 4.1 Elasticity problem for wind turbine

### 4.1.1 Governing equation

One now specialises the generic PDE in Eq. (10) for modelling elastic systems with parameter variations due to rotational components. For elastic problems, the $a(w, v; \mu)$ in variational form in Eq. (10) is defined as [24],

$$a(w, v; \mu) \equiv \int_{\Omega(\mu)} C_{ijkl}(\mu) \varepsilon_{ij}(w) \varepsilon_{kl}(v), \tag{71}$$

where $v$ is the displacement and $\varepsilon_{ij}(v)$ is the strain tensor. We consider isotropic material in this paper. Then, the coefficients $C_{ijkl}(\mu)$ are defined as [47]

$$C_{ijkl}(\mu) = \frac{E\nu}{(1+\nu)(1-2\nu)} \delta_{ij}\delta_{kl} + \frac{E\nu}{2(1+\nu)} \left( \delta_{ik}\delta_{jl} + \delta_{il}\delta_{jk} \right) \tag{72}$$

where $\nu$ denotes the Poisson's ratio, $E$ denotes the Young's modulus, and $\delta_{ij}$ denotes the Kronecker delta. Substituting Eq.(71) into Ess. (30) and (31), one can construct a reduced elastic system in the formulation of Eq. (29). To account for rotational components, the basis functions $\tilde{b}_i^f, \widetilde{\Phi}_{p,k}$ in Eqs.(30) and (31) are computed using the AP-SCRBE method discussed in Section 3. More specifically, the basis functions $\tilde{b}_i^f$ and $\tilde{b}_{i,j,k}$ are computed on the fly using Eqs. (69) and (70), in which we can set the rotation operator $\mathcal{R}(\theta)$ for different angles. $\widetilde{\Phi}_{p,k}$ are then further computed based on $\tilde{b}_{i,j,k}$ using Eqs. (27) and (28).

### 4.1.2 Offline library

The AP-SCRBE can be applied to the system to address a family of problems related to certain parameter variations after decomposing a physical asset into archetype components. For the wind turbine system, the structure of an offshore wind turbine could be decomposed into typical components such as monopile, tower, rotor and nacelle, as shown in Figure 2. The rotor could be further decomposed into blade and hub and eventually into sub-components of blade segments. Due to the repetitive nature of the structure of the wind turbine model, many of the components are derived from the same archetype. With the same archetype components in the offline library, a series of topology variations can be explored [48-50]. One can simulate a single blade, a blade-hub combination, a wind turbine, or a wind turbine installed on a monopile. However, the geometric parameter variation due to the rotation of components cannot be considered using the original SCRBE method.

In this section, the proposed AP-SCRBE is applied to build the elastic model of a wind turbine with the variation of blade pitch angle and rotor rotation angle illustrated in Figure 3. In this problem, the rotor rotates about a horizontal axis with respect to the stationary nacelle. Each blade can also undergo pitch motion by rotating with respect to the central hub to which it connects. The angle between the hub and the nacelle is defined as the rotation angle. The angle between one blade and the hub is defined as the pitch angle. Modelling the variations of pitch angle and rotor rotation angle is crucial to evaluate the performance of the turbine and its dynamic structural response. As shown in Figure 4, four adaptive ports are built to account for the variations in rotation angles. The buffer layers for the adaptive ports are highlighted and indicated by the arrows. The AP-SCRBE allows each blade to rotate at different pitch angles dependently, although the pitch angles of all blades are usually kept the same in practical operation scenarios. To model the wind turbine structure more realistically, both the blades and tower are modelled using hollow shell structures. Shear webs are also modelled inside the blades, as shown in Figure 5.

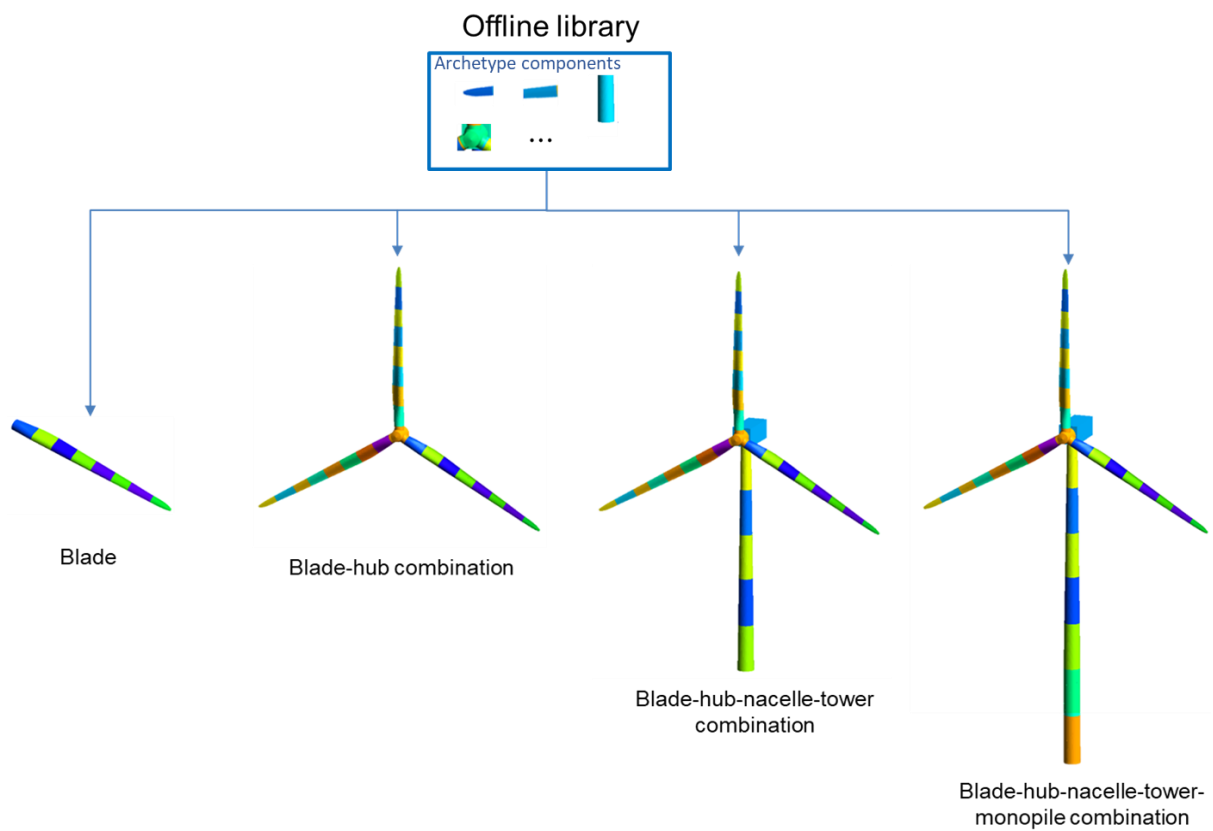

Figure 2: Topology variations using the offline archetype component library.

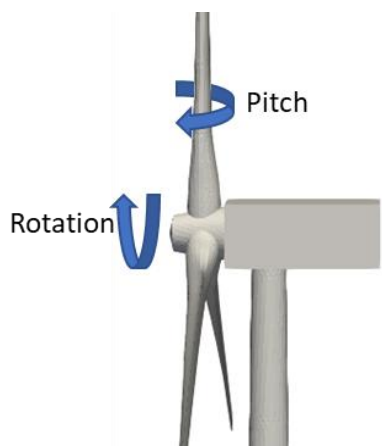

Figure 3: Wind turbine with rotational parts (e.g., blade pitch rotation and rotor rotation)

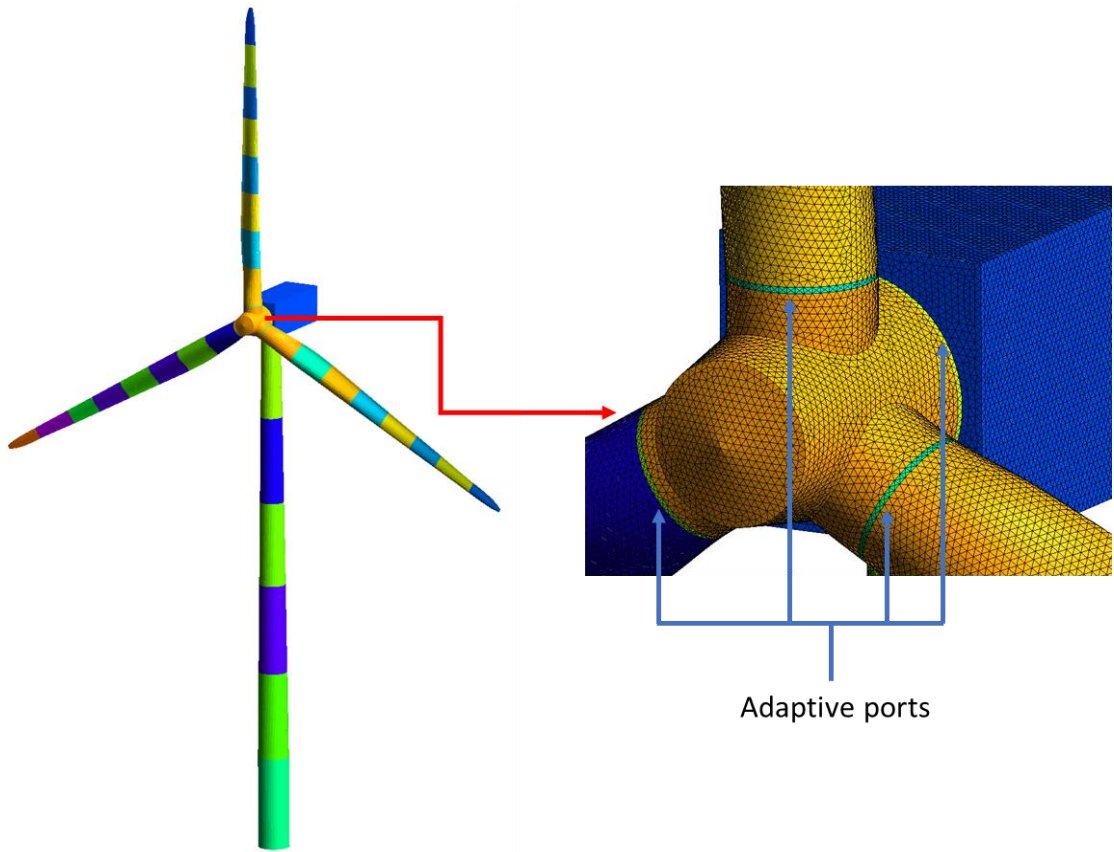

Figure 4: Three adaptive ports for modelling the pitch of the three blades and rotation of the rotor. The adaptive ports (indicated by the arrows) connect the blades to the hub and then the hub to the nacelle.

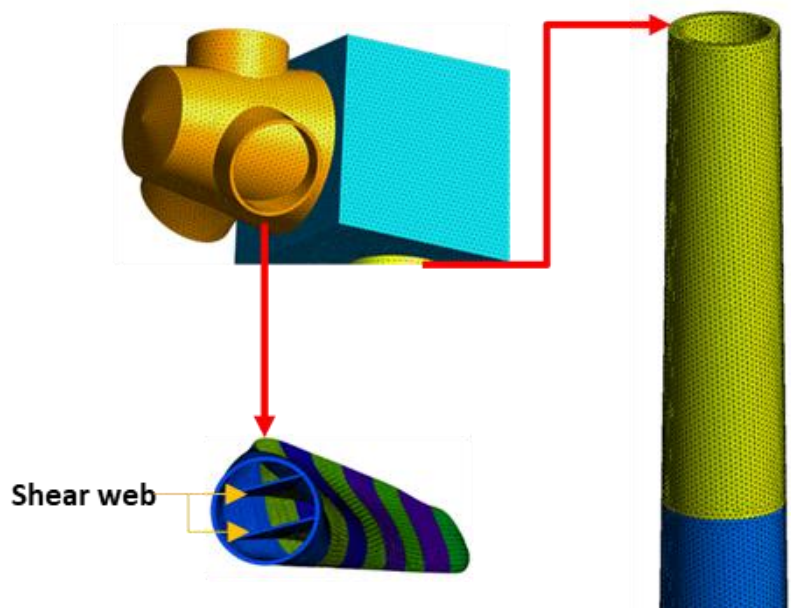

Figure 5: Internal structure of blade and tower.

## 4.2 A hub-blade combination at different pitch angles

### 4.2.1 Model description

First, a relatively simple geometry of a hub-blade combination is considered to show the numerical procedure and performance of the AP-SCRBE method. The hub-blade combination is made of one hub component (component 1) and eight blade segment components (components $2^m$, 3 to 9) instantiated from the offline archetype library shown in Figure 6. To accommodate the change of blade pitch angle, an adaptive port between the hub and blade is built, which introduces one dynamically constructed buffer layer $2^b$. When the pitch angle of the blade is changed, components $2^m$, 3 to 9 are rotated about the pitch axis, while component 1 remains stationary. Component $2^b$ is dynamically re-created on the fly and combined with the existing component $2^m$ to form a new component 2. Then, the basis functions associated with the new component 2 are computed using the AP-SCRBE method described in Section 3.

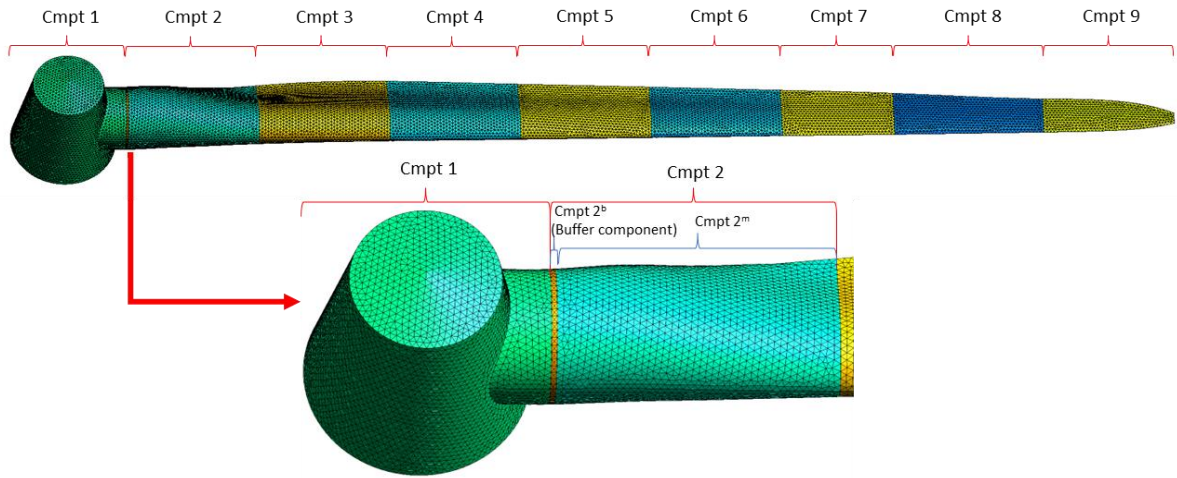

Figure 6: Computational grid of hub-blade combination.

### 4.2.2 Blade deformation at different pitch angles

Five different blade pitch angles (pitch angles = -90°, -45°, 0°, 45° and 90°) are considered as shown in Figure 7 (a). The instantiated components with underlying Fem-discretised meshes for each pitch angle are shown in Figure 7 (b) to Figure 7 (f). By applying the AP-SCRBE method, only the buffer layers highlighted in red are generated on the fly. The rest of the components are instantiated from the offline library. After that, one can rapidly simulate the deformation of the hub-blade combination under external loads. In the simulations, the Young's Modulus and density of the shear web are set to 100GPa and 1200Kg/m$^3$, while the Young's Modulus and density of the rest parts are set to 30GPa and 1800Kg/m$^3$.

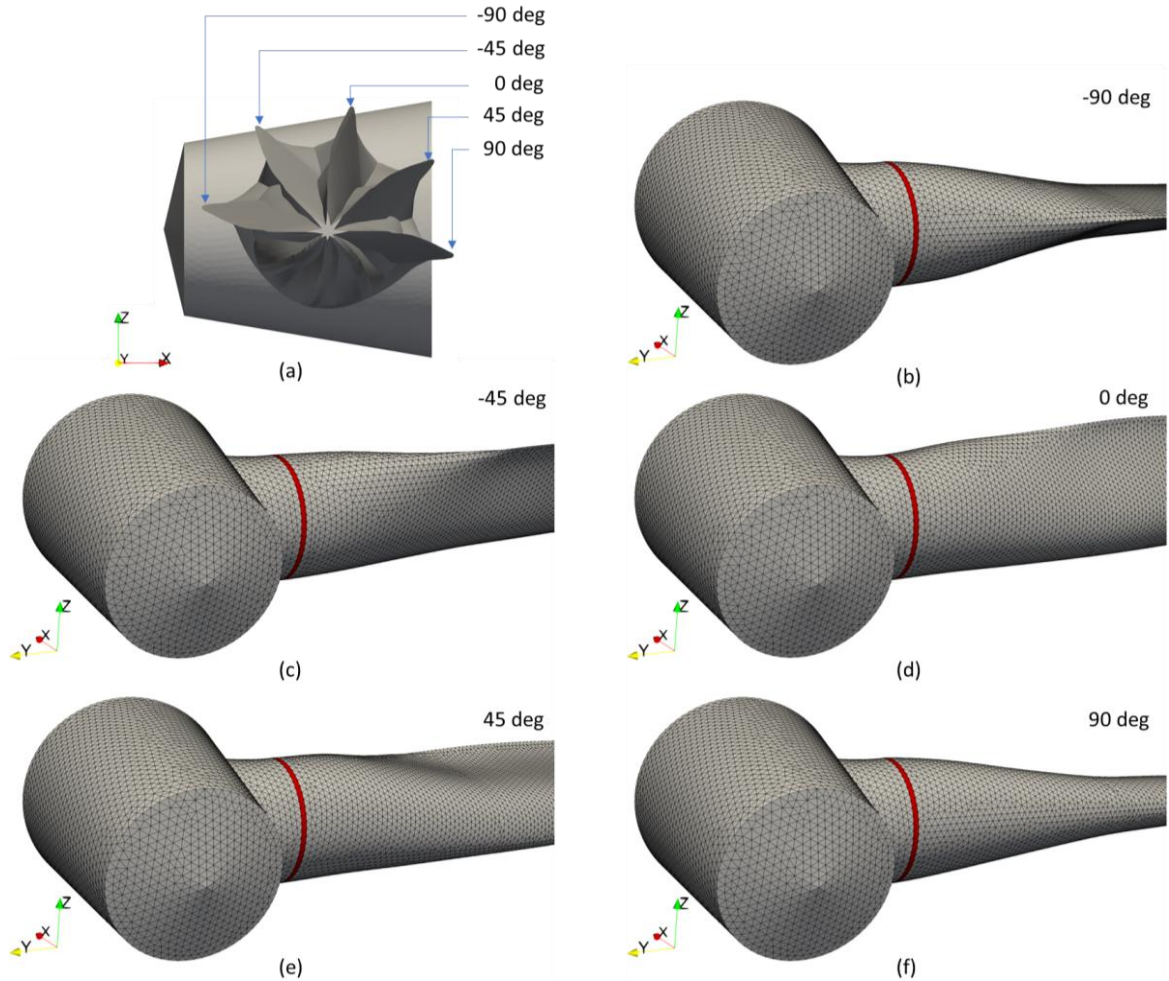

Figure 7: Visualisation of pitch rotation of one blade at different angles.

In the first test, the structural deformation is simulated under static load, i.e. the gravity force (the gravity force is along the -Z direction). First, the structural displacement is computed using both the AP-SCRBE and FEM. Then, the corresponding Von-Mises stress contours are compared for each pitch angle in Figure 8. When the pitch angle is at 0°, the high-stress region is near the edge of the blade. When the pitch angle is shifted to 45° and 90°, the high-stress region moves to the central region of the blade. Although the locations of high-stress regions are quite different at different blade pitch angles, the numerical results of AP-SCRBE consistently have good agreement with the high-fidelity FEM results in all cases. More comparisons are made in line plots by slicing the blade using a plane at X = 0. Figure 9 shows the stress on both upper and lower surfaces along the cutting line with respect to the radius (i.e. the distance to the centre of the hub). When the pitch angle is at 0°, the maximum stress occurs around a radius of 13m. When the pitch angles are at -90°, -45°, 45° and 90°, the maximum stress occurs around a radius of 30m. As the pitch angle is changed to -90° or 90° from 0°, the overall magnitude of stress gradually increases. It should be noted that the results of the AP-SCRBE and FEM are nearly identical in all cases, as shown in Figure 9. The Relative Root Mean Square Error (RRMSE) of the Von-Mises stress,

$$\text{RRMSE} = \frac{\left\|\sigma_{\text{Von-Mises}}^{\text{AP-SCRBE}} - \sigma_{\text{Von-Mises}}^{\text{FEM}}\right\|}{\left\|\sigma_{\text{Von-Mises}}^{\text{FEM}}\right\|} \tag{73}$$

and the Relative Error (RE) of the maximum Von-Mises stress,

$$\text{RE}_{\max} = \left|\frac{\max\left(\sigma_{\text{Von-Mises}}^{\text{AP-SCRBE}}\right) - \max\left(\sigma_{\text{Von-Mises}}^{\text{FEM}}\right)}{\max\left(\sigma_{\text{Von-Mises}}^{\text{FEM}}\right)}\right| \tag{74}$$

are shown in Figure 10. For all pitch angles, the RRMSE is less than 0.3%. The RE of the maximum von Mises stress is even lower, which is less than 0.01%.

In the AP-SCRBE method, a reduced elastic system is formed for each blade pitch angle. The computational space associated with the reduced elastic system is much smaller than the full computational space induced by the FEM discretisation. Table 2 shows the degrees of freedom (DoFs) of the computational spaces formed by the high-fidelity FEM and AP-SCRBE. The total number of DoFs of the computational space formed by the AP-SCRBE is 194, which is only 0.15% of that of the high-fidelity FEM. The AP-SCRBE method has a very high computational efficiency, benefiting from the reduced computational space. Table 3 summarises the computational time for each pitch angle. Since the adaptive port technique is applied, a buffer layer is generated, and related basis functions are computed on the fly for the AP-SCRBE method. The corresponding computational time is included in the online evaluation time. For FEM, the buffer mesh is also used to accommodate the pitch rotation and compute solutions in full Fem-discretised computational space. Compared to the high-fidelity FEM method, the average speed-up factor of the AP-SCRBE method is 198.

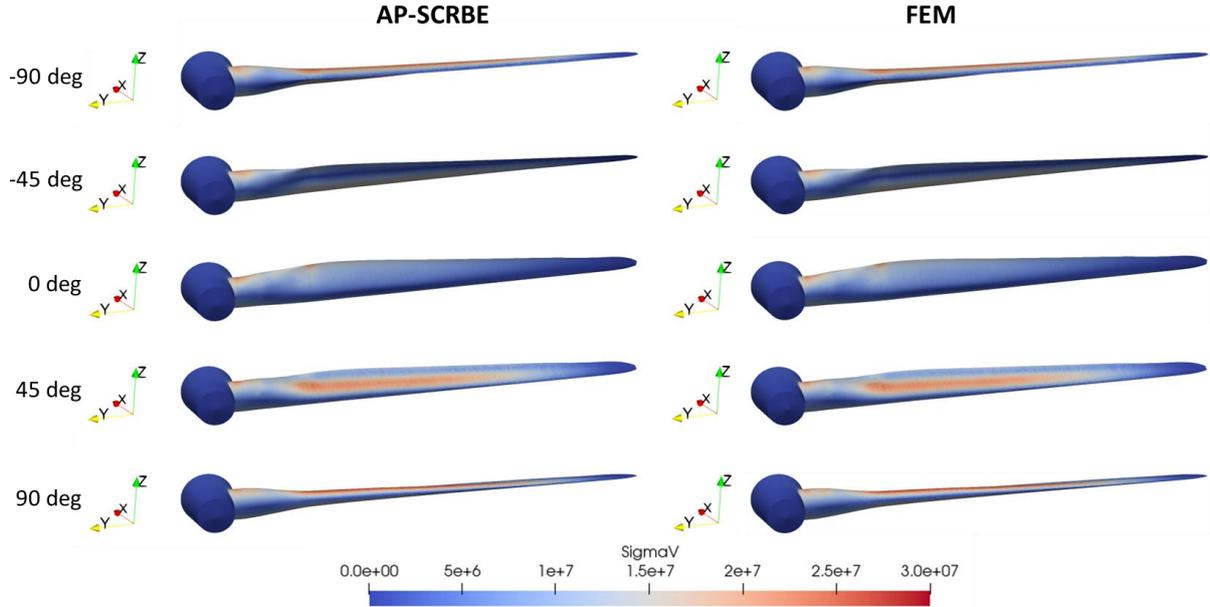

Figure 8: von Mises stress on blade surfaces at different angles.

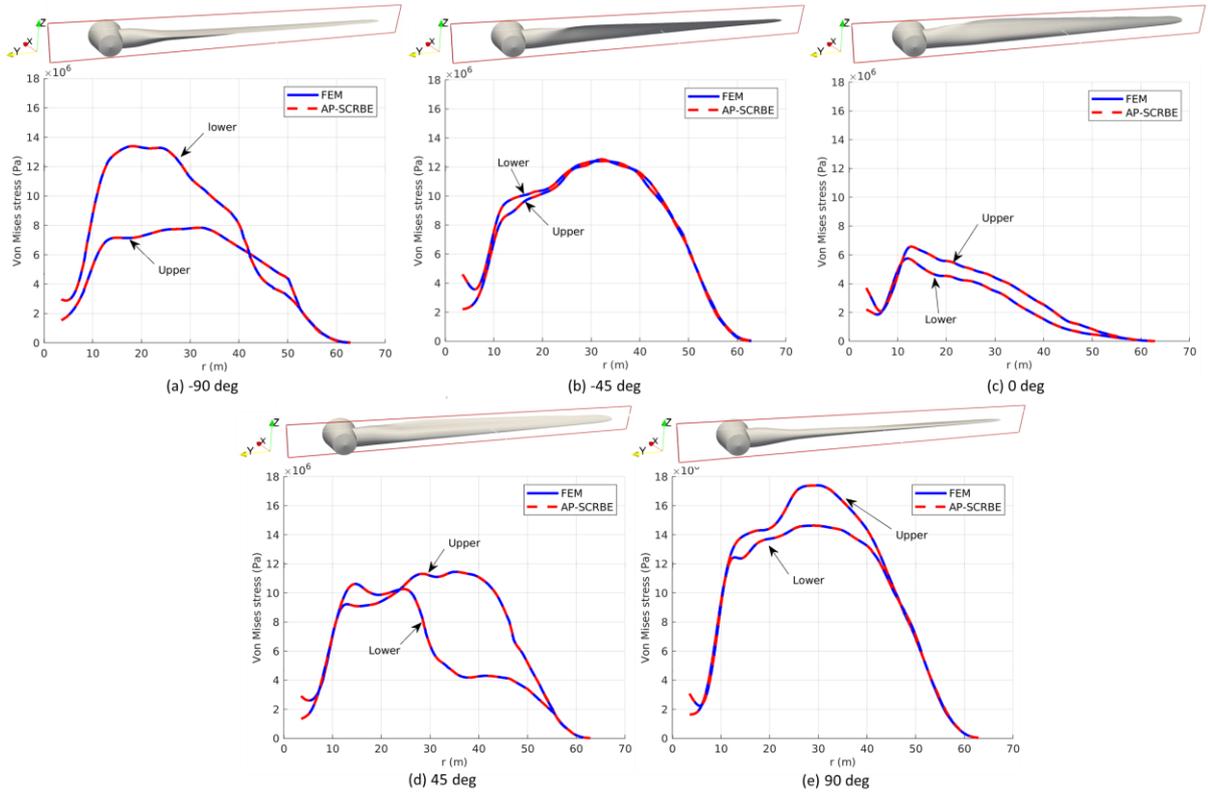

Figure 9: von Mises stress on blade surfaces along the cutting line with respect to the radius at different angles.

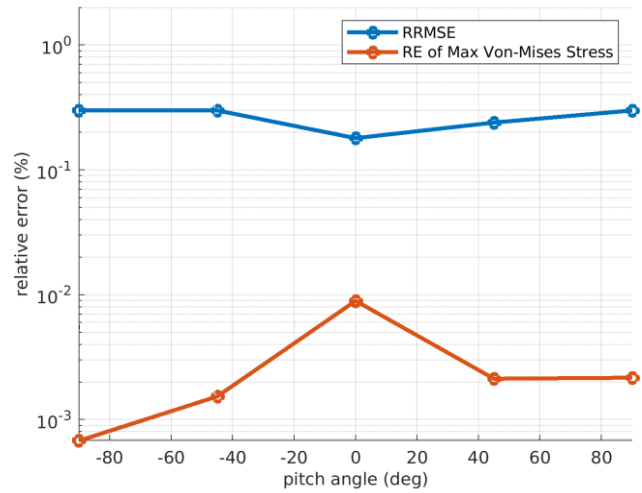

Figure 10: Error of von Mises stress on the blade at different pitch angles.

Table 2: Model sizes of the high-fidelity FEM and the AP-SCRBE.

|  | High-fidelity FEM | AP-SCRBE |
|---|---|---|
| Total number of DoFs | 127,518 | 194 |
| Percentage | 100% | 0.15% |

Table 3: Computational time of the high-fidelity FEM and the AP-SCRBE.

| | Pitch angle (°) | Online evaluation time (Sec) | | | | Average time | Speed-up factor |
|---|---|---|---|---|---|---|---|
| | | Buffer mesh generation | Bubble function | Solution Computation | Total time | | |
| FEM | -90 | $1.98 \times 10^{-3}$ | N.A. | 1.94 | 1.94 | 1.99 | N.A. |
| | -45 | $2.16 \times 10^{-3}$ | N.A. | 1.99 | 1.99 | | |
| | 0 | $2.01 \times 10^{-3}$ | N.A. | 2.05 | 2.05 | | |
| | 45 | $2.07 \times 10^{-3}$ | N.A. | 2.01 | 2.01 | | |
| | 90 | $2.10 \times 10^{-3}$ | N.A. | 1.96 | 1.96 | | |
| AP-SCRBE | -90 | $1.98 \times 10^{-3}$ | $7.45 \times 10^{-3}$ | $5.98 \times 10^{-4}$ | $10.03 \times 10^{-3}$ | $10.07 \times 10^{-3}$ | 198 |
| | -45 | $2.16 \times 10^{-3}$ | $7.28 \times 10^{-3}$ | $5.31 \times 10^{-4}$ | $9.97 \times 10^{-3}$ | | |
| | 0 | $2.01 \times 10^{-3}$ | $7.62 \times 10^{-3}$ | $5.58 \times 10^{-4}$ | $10.19 \times 10^{-3}$ | | |
| | 45 | $2.07 \times 10^{-3}$ | $7.50 \times 10^{-3}$ | $5.30 \times 10^{-4}$ | $10.10 \times 10^{-3}$ | | |
| | 90 | $2.10 \times 10^{-3}$ | $7.41 \times 10^{-3}$ | $5.32 \times 10^{-4}$ | $10.04 \times 10^{-3}$ | | |

## 4.3 Wind turbine at different rotor rotation angles

### 4.3.1 Model description

Next, the AP-SCRBE method is applied to build an elastic model of the NREL 5MW wind turbine at different rotor rotation angles. Figure 11 shows the model configuration of the wind turbine when the rotor rotation angle is at 0°. In this simulation, the rotor rotates with respect to the stationary nacelle and the tower along the horizontal axis at a given blade pitch angle. The components inside the blade, hub, nacelle and tower are all instantiated from the offline archetype library. The pitch of the blades is handled with buffer layers (highlighted in green) using the strategy described in Section 4.2 . The pitch angle is set to 0° at the beginning and fixed for the rest of the simulation. To accommodate the rotation of the turbine rotor, a new hub component with an adaptive port is created online by generating a buffer layer (highlighted using the red colour) for any rotation angle.

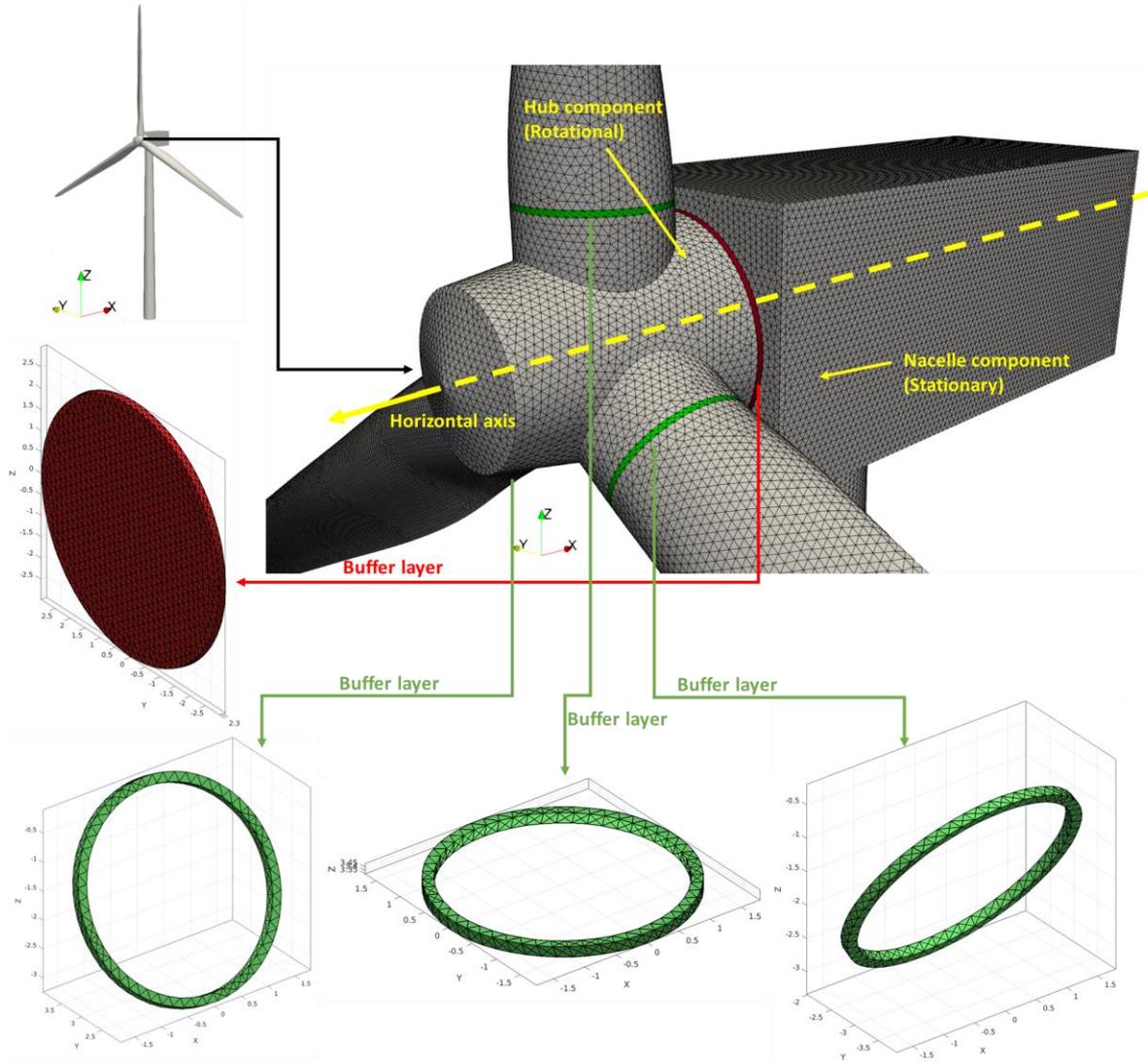

Figure 11: Configuration of the wind turbine with buffer layers for the AP-SCRBE.

### 4.3.2 Wind turbine deformation at different rotor rotation angles

Figure 12 shows the wind turbine at four different rotor rotation angles (rotation angles = 0°, 30°, 60° and 90°) considered in the simulations. At each rotation angle, only the buffer layers highlighted in red and green colours are generated on the fly. The rest of the components are instantiated from the offline library. For convenience, we use component $1^m$ and component $1^b$ to denote the instantiated hub and buffer layers, respectively. When the rotor rotation angle is changed, component $1^b$ is dynamically re-created on the fly and combined with the instantiated component $1^m$ to form a new component 1. Then, the basis functions are computed to construct a reduced elastic system. In the simulations, the Young's Modulus and density of the shear web are set to 100GPa and 1200Kg/m$^3$. The Young's Modulus and density of the other parts in the blades are set to 30GPa and 1800Kg/m$^3$. The Young's Modulus and density of the supporting tower are set to 200GPa and 8000Kg/m$^3$.

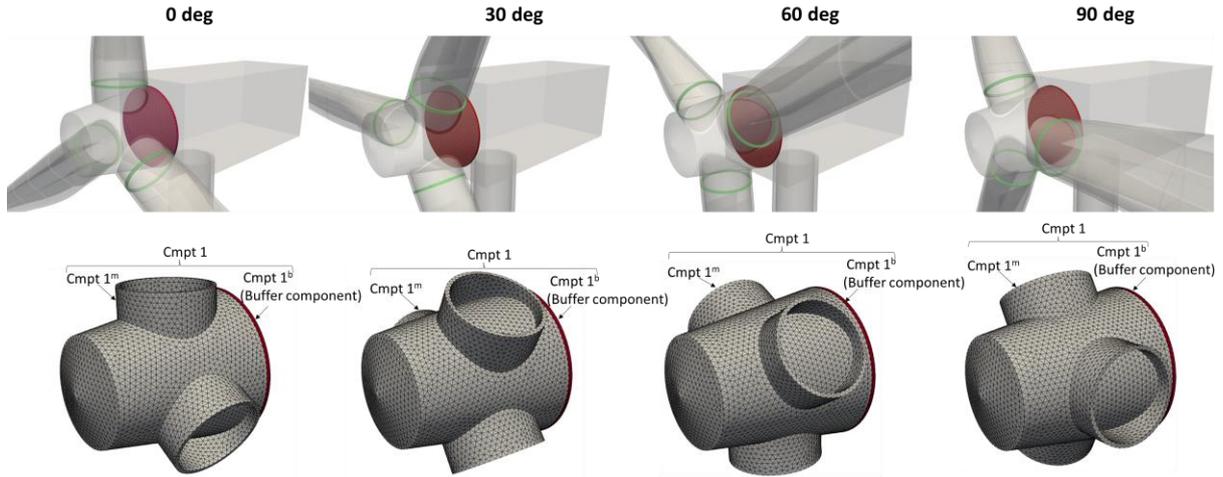

Figure 12: Wind turbine and the buffer layers at different rotation angles.

The operating condition of the rotor rotating at 12 rpm and a blade pitch angle of 0° is considered. The wind direction is normal to the front of the wind turbine at a speed of 12 m/s. The external loads include the wind, gravity, and centrifugal forces on the rotor. The wind load is computed using the computational fluid dynamics simulation [49] under the quasi-static Multiple Reference Frame (MRF) [51, 52]. The pressure contour of the wind load is shown in Figure 13 at different rotation angles. The weight of the gearbox is assumed to be 70 tons, and the load is uniformly distributed over the nacelle.

The structural displacement is computed using both the AP-SCRBE and FEM. Then, the corresponding von Mises stress contours are compared for each rotation angle in Figure 14. For blade 1, as the rotation angle changes from 0° to 90°, the high-stress region starts concentrating around the trailing edge of blade 1 at approximately 1/5 span. For blade 2, the high-stress region appears on the front surface of the blade at approximately 1/5 span. As the rotation angle changes from 0° to 60°, the overall magnitude of stress gradually decreases. For blade 3, the high-stress region appears near the leading edge of the blade. As the rotation angle changes from 30° to 90°, the magnitude of stress gradually decreases. In all cases, the stress distributions computed using the AP-SCRBE are almost identical to the numerical results computed using the high-fidelity FEM. Figure 15 shows the stress on both front and back surfaces along a cutting line at the middle plane of blade 1 with respect to the radius (distance to the centre of the hub). The result lines of the AP-SCRBE and FEM nearly overlap with each other in all cases. Figure 16 shows the RRMSE of the von Mises stress and the RE of the maximum von Mises stress, which are defined in Eq.(73) and Eq.(74). For all rotation angles, the RRMSE is less than 0.3%. The RE of the maximum von Mises stress is even lower, which is less than 0.01%.

Table 4 shows the DoFs of the computational spaces formed by the high-fidelity FEM and AP-SCRBE. The total number of DoFs of the computational space formed by the AP-SCRBE is 909, which is only 0.11% of the computational space of the high-fidelity FEM. Table 5 lists out the breakdown of the online computational time for each rotation angle. In this simulation, the wind turbine model consists of 32 components and 31 ports, of which 4 ports are adaptive ports. The model involves more components and, therefore, is more complex. Consequently, the time consumption for buffer layer generation, basis function computation, and solution computation increases compared to the simulations in Section 4.2 . A higher speed-up factor can also be expected for problems with more complexity. Compared to the high-fidelity FEM method, the average speed-up factor of the AP-SCRBE method increases to nearly three orders.

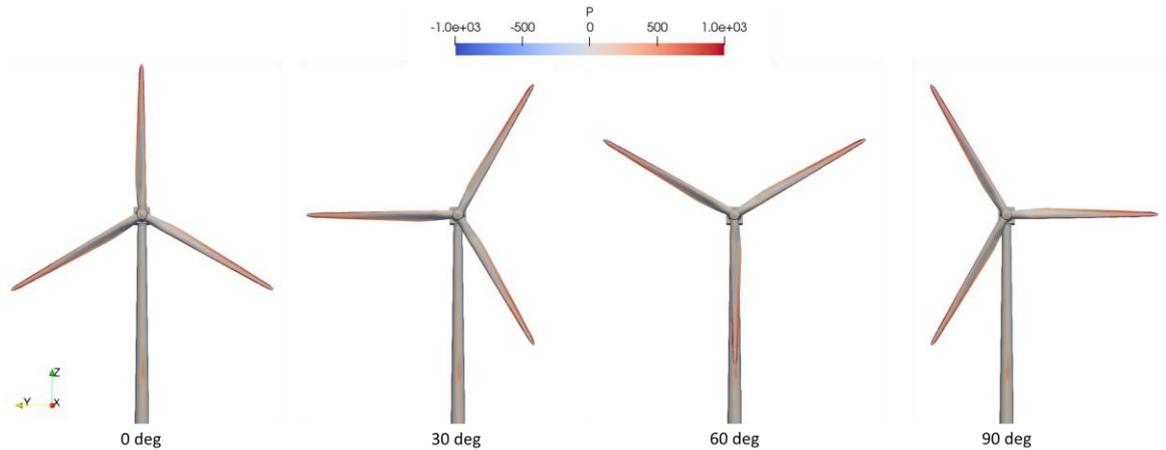

Figure 13: Pressure contour of the wind load on the turbine at each rotation angle.

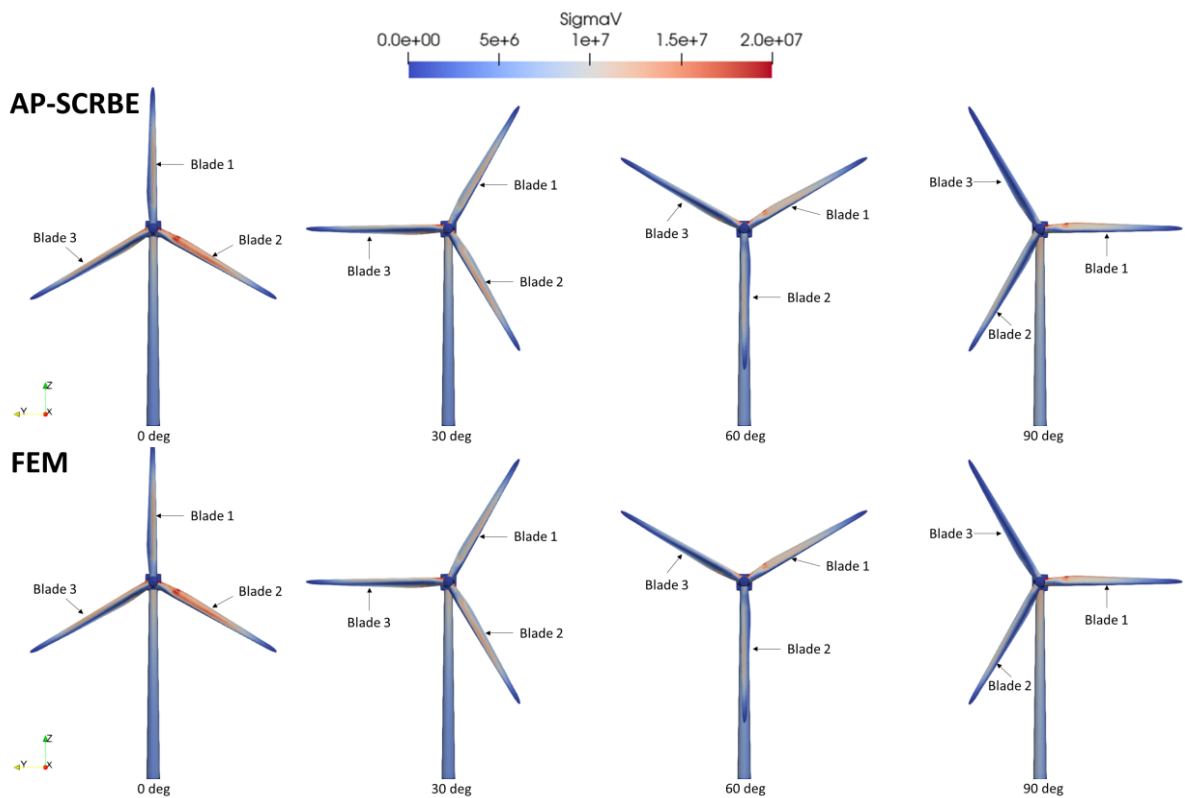

Figure 14: von Mises stress contours on the turbine at each rotor angle.

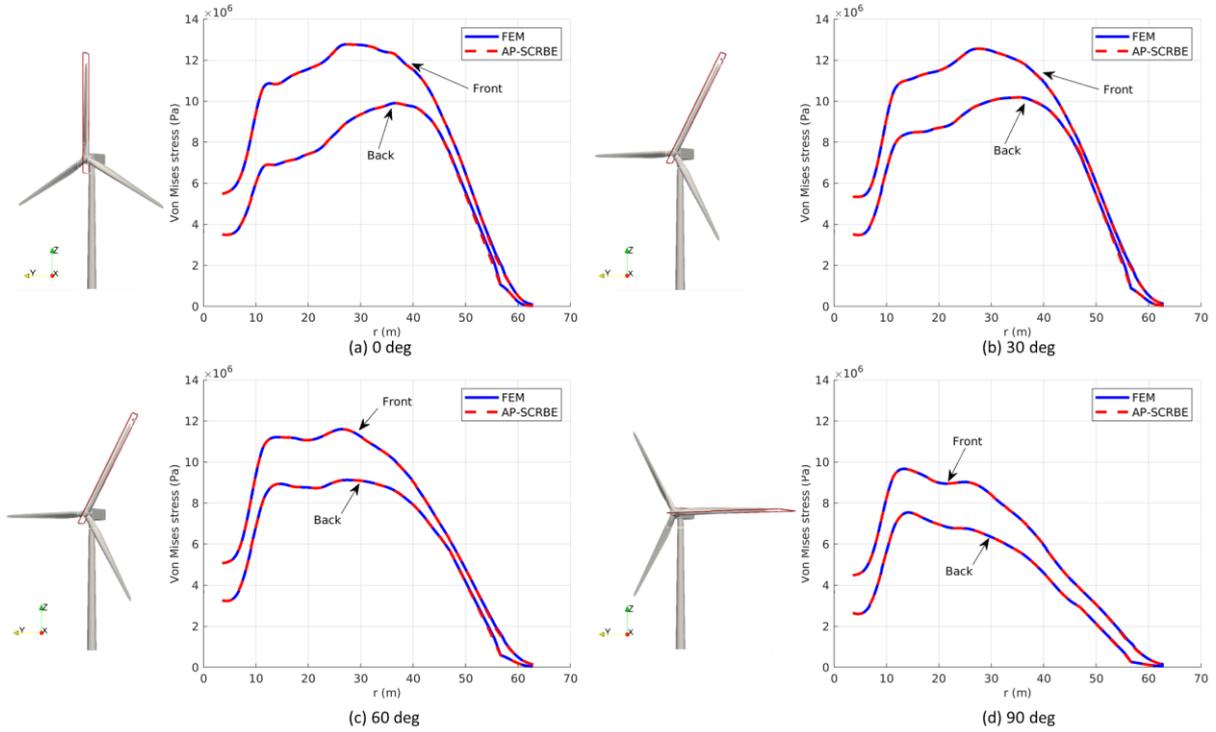

Figure 15: von Mises stress on front and back surfaces along the cutting line at the middle plane of blade 1.

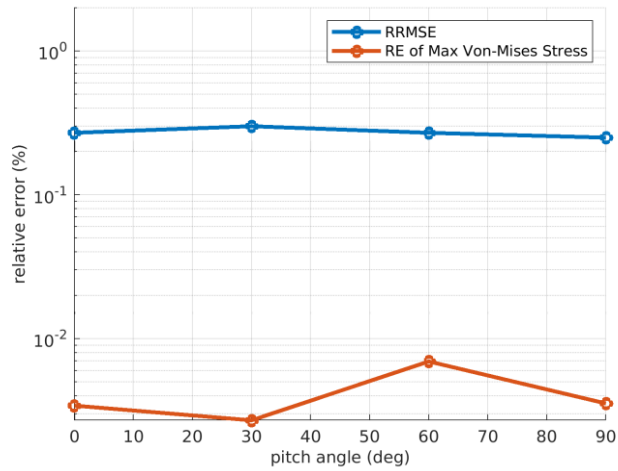

Figure 16: Error of von Mises stress on blade 1 at different rotor angles.

Table 4: Model sizes of the high-fidelity FEM and the AP-SCRBE.

|  | High-fidelity FEM | AP-SCRBE |
| --- | --- | --- |
| Total number of DoFs | 803,298 | 909 |
| Percentage | 100% | 0.11% |

Table 5: Computational time of the high-fidelity FEM and the AP-SCRBE.

| | Rotation angle (°) | Online evaluation time (Sec) | | | | Average time | Speed-up factor |
|---|---|---|---|---|---|---|---|
| | | Buffer mesh generation | Bubble function | Solution Computation | Total time | | |
| FEM | 0 | $1.36 \times 10^{-2}$ | N.A. | 54.30 | 54.31 | 53.54 | N.A. |
| | 30 | $1.28 \times 10^{-2}$ | N.A. | 53.10 | 53.11 | | |
| | 60 | $1.24 \times 10^{-2}$ | N.A. | 55.03 | 55.15 | | |
| | 90 | $1.25 \times 10^{-2}$ | N.A. | 51.46 | 51.58 | | |
| AP-SCRBE | 0 | $1.36 \times 10^{-2}$ | $3.50 \times 10^{-2}$ | $1.73 \times 10^{-2}$ | $6.59 \times 10^{-2}$ | $6.45 \times 10^{-2}$ | 831 |
| | 30 | $1.28 \times 10^{-2}$ | $3.43 \times 10^{-2}$ | $1.82 \times 10^{-2}$ | $6.53 \times 10^{-2}$ | | |
| | 60 | $1.24 \times 10^{-2}$ | $3.37 \times 10^{-2}$ | $1.83 \times 10^{-2}$ | $6.44 \times 10^{-2}$ | | |
| | 90 | $1.25 \times 10^{-2}$ | $3.38 \times 10^{-2}$ | $1.59 \times 10^{-2}$ | $6.22 \times 10^{-2}$ | | |

# 5 Conclusions

This paper proposed an AP technique to extend CMS approaches for modelling large-scale parametric systems with rotational parts. In particular, a variant of CMS approaches, the SCRBE method, is used to demonstrate the capability of the AP technique in synthesising both stationary and rotational components. The new version of the SCRBE method, which has been improved by the AP technique, is called AP-SCRBE. Similarly to the original SCRBE, the AP-SCRBE is a component-based approach, which is very suitable for complex systems featuring repetitive geometrical structures. On top of that, the AP-SCRBE extends the component-based approach to parametric systems with rotational parts. The key to the AP-SCRBE is the adaptive port technique. This technique resolves the discretisation incompatibility on connecting ports of the basic SCRBE by introducing a buffer layer. More importantly, one can use this technique to rapidly compute basis functions for rotational components, so that the rotational components can be synthesised with stationary components very efficiently.

Both AP-SCRBE and SCRBE are component-based ROM approaches, which rely on the computation of bubble functions within the instantiated components. The bubble functions are affected by the neighbouring element of the components. Therefore, the accuracy of the resulting ROM depends on the locality assumption, which typically holds true for diffusion-dominated systems. The accuracy could potentially drop for convection-dominated systems, especially those with high nonlinearity. In this work, the AP-SCRBE method is applied to the elastic modelling of a wind turbine within the context of rotational-stationary component synthesis, which is a diffusion-dominated problem. The AP-SCRBE can accurately and efficiently model the wind turbine with the pitch rotation of the blades and the axial rotation of the rotor under quasi-static loads. The results from the AP-SCRBE are almost identical to those of the high-fidelity FEM, with the RRMSE being less than 0.3%. The numerical examples also show that the AP-SCRBE can achieve a two-order-of-magnitude speed-up in computation at the cost of a minor drop in accuracy compared to the high-fidelity FEM. Beyond that, the AP-SCRBE is not limited to problems with rotational parts. It can also potentially be used for other systems with more complex part movements.

# Acknowledgements

The work presented in this paper is a result of research conducted through the Enabling Future Systems for Offshore Wind Resources (ENFORCE) program, supported by A*STAR under its RIE 2025 Industry Alignment Fund (Grant No.: M23M4a0067).

# Statements and Declarations

The authors declare that they have no known competing financial interests or personal relationships that could have appeared to influence the work reported in this paper.